# The vagus nerve regulates immunometabolic homeostasis in the ovine fetus near term: impact on terminal ileum

Mingju Cao[1], Shikha Kuthiala[1], Keven Jason Jean[1], Hai Lun Liu[1], Marc Courchesne[2], Karen Nygard[2], Patrick Burns[3], André Desrochers[3], Gilles Fecteau[3], Christophe Faure[4], and Martin G. Frasch[1,5,6]

[1] Dept. of OBGYN and Dept. of Neurosciences, CHU Ste-Justine Research Centre, Université de Montreal, Montréal, QC, Canada
[2] Biotron Microscopy, Western University, London, ON, Canada
[3] Clinical Sciences, CHUV, Université de Montreal, St-Hyacinthe, QC, Canada
[4] Dept. of Pediatrics, CHU Ste-Justine Research Centre, Université de Montreal, Montréal, QC, Canada
[5] Centre de recherche en reproduction animale, l'Université de Montréal, St-Hyacinthe, QC, Canada
[6] Dept. of OBGYN and Center on Human Development and Disability, University of Washington School of Medicine, CHDD, Seattle, WA, USA

**Corresponding author:**
Martin G. Frasch
Department of Obstetrics and Gynecology
University of Washington
1959 NE Pacific St
Box 356460
Seattle, WA 98195
Phone: +1-206-543-5892
Fax: +1-206-543-3915
Email: mfrasch@uw.edu

## Key points

- Brain senses and regulates body's glucose levels and the inflammatory status via the vagus nerve, but the precise contribution and significance in the developing fetus are not understood.

- In a large prospective cohort of chronically instrumented fetal sheep, the best model of fetal system's physiology, we demonstrate that a complete removal of brain-body connection via the vagus nerve results in a sustained increase in glucose levels and an intermittent hyperinsulinemia.

- Under conditions of moderate fetal inflammation, this response is related to higher levels of gut inflammation.

- The efferent vagus nerve stimulation reduces the systemic inflammatory response as well as restores both the levels of glucose and terminal ileum inflammation, but not the insulin levels.

- Moreover, our findings reveal a novel regulatory, hormetic, role of the vagus nerve in the immunometabolic response to endotoxin in near-term fetuses.

# Abstract

BACKGROUND. Glucosensing elements are widely distributed throughout the body and relay information about circulating glucose levels to the brain via the vagus nerve. However, while anatomical wiring has been established, little is known about the physiological role of the vagus nerve in glucosensing. The contribution of the vagus nerve to inflammation in the fetus is poorly understood. Increased glucose levels and inflammation act synergistically in causing organ injury, but their interplay remains incompletely understood. We hypothesized that vagotomy (Vx) will trigger rise in systemic glucose levels and this will be enhanced during systemic and organ-specific inflammation. Efferent vagus nerve stimulation (VNS) should reverse this phenotype.

METHODS. Near-term fetal sheep (n=57) were surgically prepared with vascular catheters and ECG electrodes as control and treatment groups (lipopolysaccharide (LPS), Vx+LPS, Vx+LPS+selective efferent VNS). The experiment was started 72 hours postoperatively to allow for post-surgical recovery. Inflammation was induced with LPS bolus intravenously (LPS group, 400 ng/fetus/day for 2 days; N=23). Vx+LPS group consisted of eleven animals: a bilateral cervical vagotomy was performed during surgery; of these n=5 received double the LPS dose, LPS800. Vx+LPS+efferent VNS group received cervical VNS probes bilaterally distal from Vx in eight animals. Efferent VNS was administered for 20 minutes on days 1 and 2 +/10 minutes around LPS bolus. Fetal arterial blood samples were drawn on each postoperative day of recovery (-72 h, -48 h, and -24 h) as well as at baseline and seven selected time points (+3-48 h) to profile inflammation (ELISA IL-6, pg/mL), insulin (ELISA), blood gas and metabolism (glucose). At 54 h post LPS, a necropsy was performed; terminal ileum macrophages' CD11c (M1 phenotype) immunofluorescence was quantified to detect inflammation. Results are reported for $p<0.05$ and for Spearman $R2>0.1$. Results are presented as median(IQR).

RESULTS. Across the treatment groups, blood gas and cardiovascular changes indicated a mild septicemia. At 3 h, in the LPS group IL-6 peaked; that peak was decreased in Vx+LPS400 and doubled in Vx+LPS800 group; the efferent VNS sped up the reduction of the inflammatory response profile over 54 h. M1 macrophage activity was increased in the LPS and Vx+LPS800 groups only. Glucose and insulin levels in the Vx+LPS group were respectively 1.3-fold and 2.3-fold higher vs. control at 3 h, and the efferent VNS normalized glucose levels.

CONCLUSIONS. Complete withdrawal of vagal innervation results in a 72h delayed onset of sustained increase in glucose for at least 54h and intermittent hyperinsulinemia. Under conditions of moderate fetal inflammation, this is related to higher levels of gut inflammation; the efferent VNS reduces the systemic inflammatory response as well as restores both the levels of glucose and terminal ileum inflammation, but not the insulin levels. Our findings reveal a novel regulatory, hormetic, role of the vagus nerve in the immunometabolic response to endotoxin in near-term fetuses.

## Visual abstract

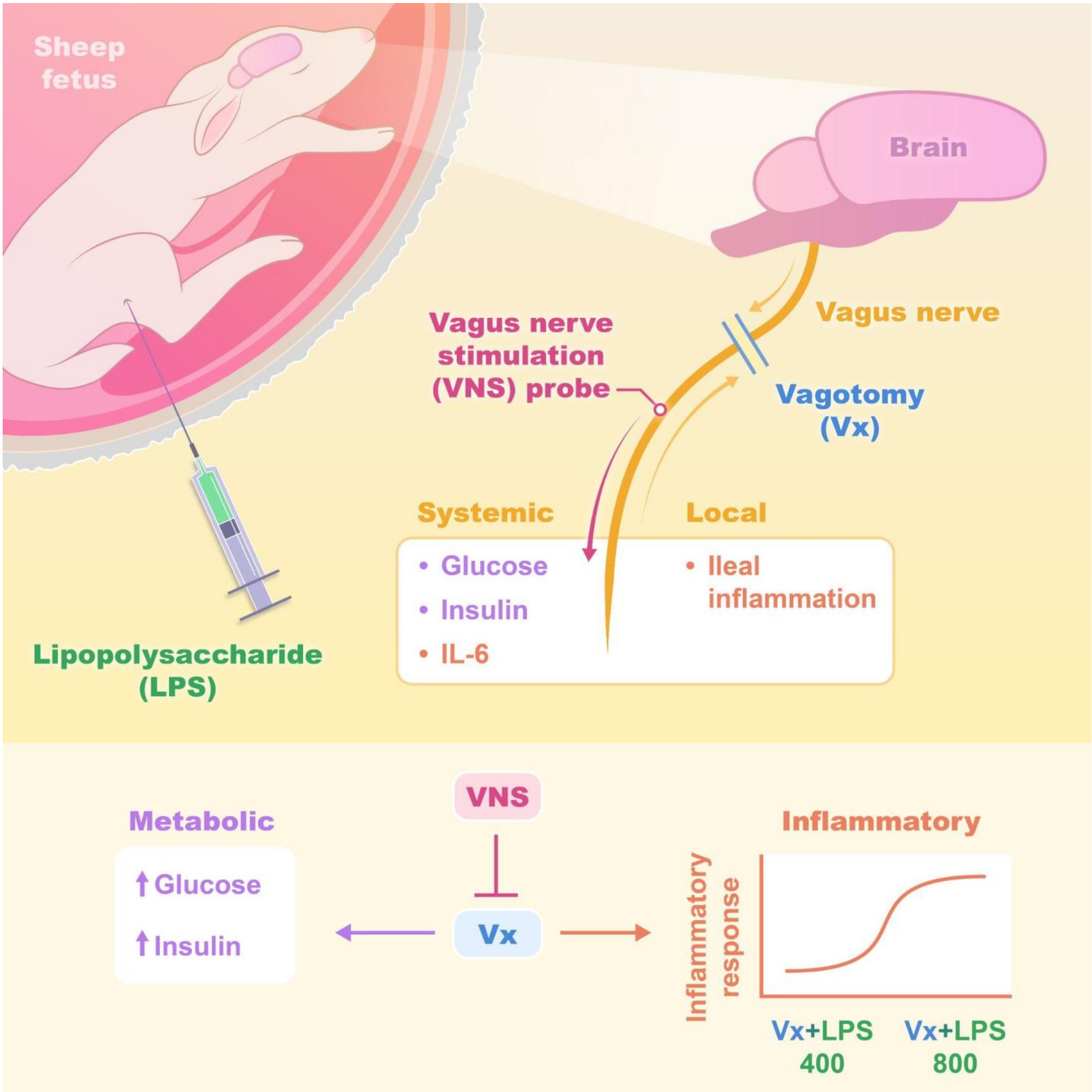

## Introduction

The fetal inflammatory response is an important independent contributor to antenatal and perinatal brain injury.(Hagberg et al., 2015; Rees and Inder, 2005; Wang et al., 2006) Via the cholinergic anti-inflammatory pathway, the vagus nerve attenuates the inflammatory response to endotoxin.(Borovikova et al., 2000) The fetal cholinergic anti-inflammatory pathway is active in pre- and near-term ovine fetuses.(Frasch et al., 2016; Garzoni et al., 2013, 2014; Kwan et al., 2016; Liu et al., 2016a) Using this important model of human fetal development, we also established the approach to manipulate fetal vagus nerve in the chronically instrumented unanesthetized fetal sheep near term.(Castel et al., 2021a; Morrison et al., 2018)

The role of glycemic control in this physiological relationship is not well understood with evidence accruing that hyperglycemia exacerbates the inflammation induced by endotoxin.(Dhillon et al., 2018) Similar assumptions are made for the etiology of necrotizing enterocolitis of the neonate (NEC) and other major neonatal causes of morbidity and mortality, especially in premature neonates.(Kao et al., 2006; van der Lugt et al., 2010),(Hall et al., 2004; LeBlanc et al., 1993)

Glucosensing elements are widely distributed throughout the body.(Grabauskas et al., 2013; Lopez-Barneo, 2003) Carotid body, gut, pancreas, and hepatic portal/mesenteric vein have been implicated in glucosensing. From these locations, the vagus nerve carries information about circulating glucose to the brain. (Levin, 2006a) However, while the anatomical wiring has been established, the physiological role of the vagus nerve in glucosensing, along with other key homeostatic processes, is only beginning to unravel with evidence coming from epidemiological and animal experimental studies.(Grabauskas et al., 2013; Herry et al., 2019; Thayer and Sternberg, 2006)

Chorioamnionitis is an antenatal proinflammatory process involving the placenta and fetal membrane, which may play a contributing role in the pathogenesis of NEC. The main manifestation of pathologic inflammation in the fetoplacental unit, chorioamnionitis, affects 20% of term pregnancies and up to 60% of preterm pregnancies.(Gotsch et al., 2007; Lahra and Jeffery, 2004) Both symptomatic and asymptomatic chorioamnionitis are associated with an increased risk of NEC.(Garzoni et al., 2013) Prevention of NEC in neonates remains a major clinical challenge and a better understanding of the interplay between the synergistic contributing mechanisms of inflammation and rising glucose levels is needed.(Joseph et al., 2019)

Insulin treatment does not provide effective and safe glycemic control in at-risk neonates. Better therapeutic modalities for glycemic control are needed. Modulation of vagus nerve activity may provide one such modality, but the first step must be a clearer elucidation of the role of the vagus nerve in the etiology of fetal inflammatory response and changes in glucose homeostasis. Advancing this understanding was the goal of the present study.

First, in humans, Thayer and Sternberg (2006) reviewed the evidence from four studies in ~18,000 healthy subjects and subjects with diabetes type 2 that decreased vagal function assessed via its proxy, heart rate variability (HRV) measures, is associated with increased fasting glucose and hemoglobin A1c levels; this shows overall that reduced vagal activity is a hallmark of diabetes.(Liao et al., 1995; Panzer et al., 2002; Singh et al., 2000; Thayer and Fischer, 2005) Moreover, chronic cervical vagus nerve stimulation (VNS) for control of drug-resistant epilepsy increases their fasting glucose levels, albeit still within healthy limits.(Liu et al., 2020)

Second, in animals, in seeming contrast to some of the studies in human subjects, a recent report indicated that VNS in adult mice exposed to endotoxin reduces inflammation-induced hyperglycemia by inducing insulin.(Joseph et al., 2019) The chronic hyperglycemic effects of VNS, reported in humans, were also validated in a murine study, likely due to non-selective VNS which activates the afferent fibers thought responsible for this effect, in contrast to the efferent fibers which are thought to mediate the hypoglycemic effects of VNS.(Meyers et al., 2016; Stauss et al., 2018) In another adult murine model study, vagus nerve electroneurogram was shown to encode glucose levels corroborating the notion that vagus nerve carries information about glucose homeostasis.(Masi et al., 2019)

Taken together, it is evident that our knowledge about vagal glycemic control across different species and developmental stages of this control is still very limited. However, it is clear that vagus nerve exerts - selectively via its efferent and afferent pathways - powerful modulatory influence on glucose *and* inflammatory homeostatic control systems in health and disease. Evidence is also compelling that the vagus nerve's regulatory role is present already during fetal development, a period when adverse exposures are known to have powerful long-lasting reprogramming effects on postnatal health and predisposition to disease in later life.(Antonelli et al., 2022; Cerritelli et al., 2021; Morrison et al., 2018)

Consequently, we hypothesized that bilateral cervical vagotomy in a mature, near-term, fetal sheep will result in increase of systemic glucose levels and this will be correlated to a higher degree of systemic inflammation and the inflammation in the terminal ileum, NEC's locus minoris resistentiae. Selective efferent VNS will reverse these patterns.

# Methods

## Ethics Approval

Animal care followed the guidelines of the Canadian Council on Animal Care and the approval by the University of Montreal Council on Animal Care (protocol #10-Rech-1560).

## Anesthesia and surgical procedure

We reported the detailed approach including Vx and VNS elsewhere.(Burns et al., 2015; Castel et al., 2021a) Briefly, we instrumented 57 pregnant time-dated ewes at 126 days of gestation (dGA, ~0.86 gestation) with arterial, venous and amniotic catheters, fetal precordial ECG and cervical bilateral VNS electrodes; 19 animals received cervical bilateral vagotomy (Vx) during surgery of which 8 animals received efferent VNS electrodes and VNS treatment during the experiment.(Burns et al., 2015; Castel et al., 2021a)

Ovine singleton fetuses of mixed breed were surgically instrumented with sterile technique under general anesthesia (both ewe and fetus). In the case of twin pregnancy, the larger fetus was chosen based on palpating and estimating the intertemporal diameter. The total duration of the procedure was carried out in about 2 hours. Antibiotics were administered to the mother intravenously (Trimethoprim sulfadoxine 5 mg/kg) as well as to the fetus intravenously and into the amniotic cavity (ampicillin 250 mg). Amniotic fluid lost during surgery was replaced with warm saline. The catheters exteriorized through the maternal flank were secured to the back of the ewe in a plastic pouch. For the duration of the experiment, the ewe was returned to the metabolic cage, where she could stand, lie and eat ad libitum while we monitored the non-anesthetized fetus without sedating the mother. During postoperative recovery antibiotic administration was continued for 3 days. Arterial blood was sampled for evaluation of the maternal and fetal condition and catheters were flushed with heparinized saline to maintain patency.

## Experimental protocol

Postoperatively, all animals were allowed 3 days to recover before starting the experiments. On these 3 days, at 9.00 am 3 mL arterial plasma sample was taken for blood gasses and cytokine analysis. Each experiment commenced at 9.00 am with a 1 h baseline measurement followed by the respective intervention as outlined below (Fig. 1). FHR and arterial blood pressure was monitored continuously (CED, Cambridge, U.K., and NeuroLog, Digitimer, Hertfordshire, U.K). Blood and amniotic fluid samples (3 mL) were taken for arterial blood gasses, lactate, glucose and base excess (in plasma, ABL800Flex, Radiometer) and cytokines (in plasma and amniotic fluid) at the time points 0 (baseline), +1 (*i.e.*, immediately after LPS administration), +3, +6, +12, +24, +48 and +54 h (*i.e.*, before sacrifice at day 3). For the cytokine analysis, plasma was spun at 4°C (4 min, 4000g force, Eppendorf 5804R, Mississauga, ON), decanted and stored at -80°C for subsequent ELISAs. After the +54 hours (Day 3) sampling, the animals were sacrificed as reported.(Burns et al., 2015; Castel et al., 2021a) Fetal growth was assessed by body, brain, liver and maternal weights.

Lipopolysaccharide (LPS)-induced inflammation in fetal sheep is a well-established model of the human fetal inflammatory response to sepsis.(Durosier et al., 2015; Nitsos et al., 2006; Rees and Inder, 2005; Wang et al., 2006)

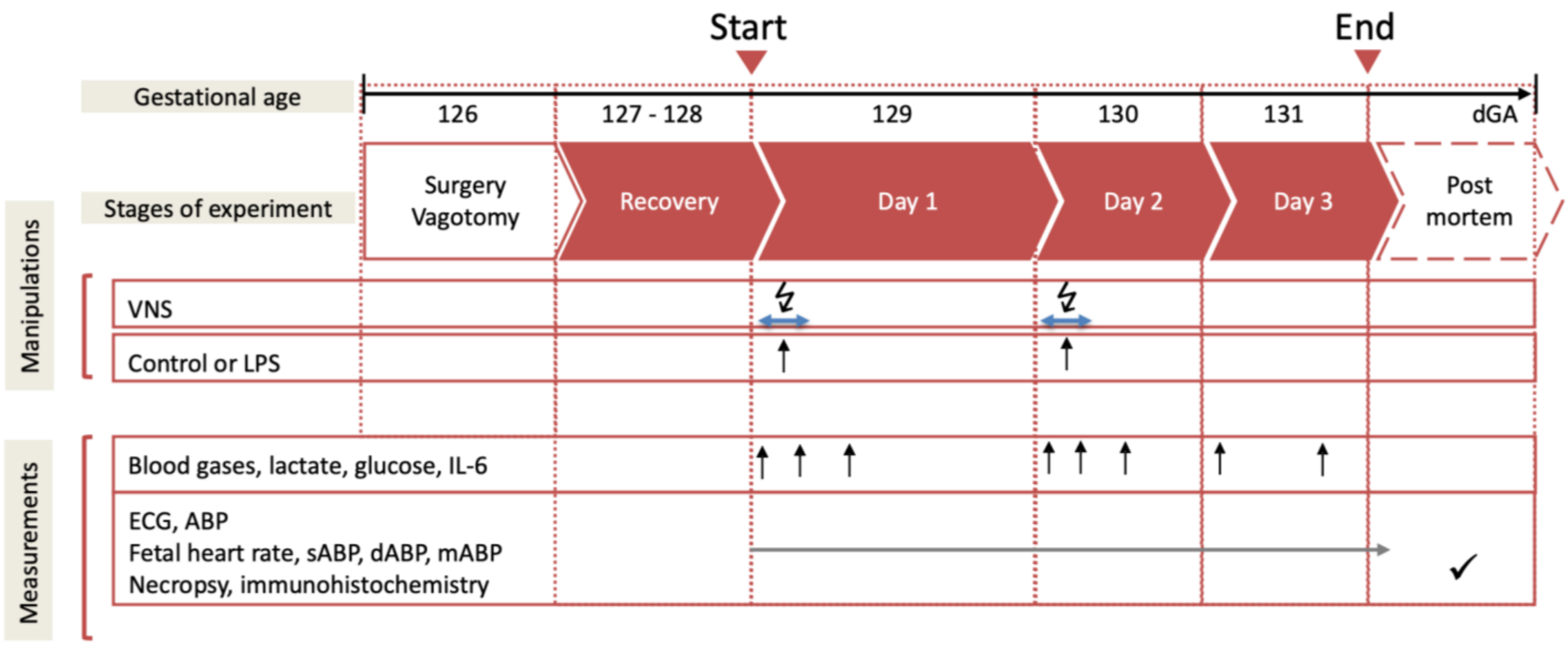

**Figure 1. Experimental design.** Bilateral cervical vagotomy (Vx) was performed during surgery in Vx+LPS group animals. At Days 1 and 2, Vx+LPS animals received LPS dose 400 or 800 ng/fetus/day. Some Vx+LPS animals also received the efferent intermittent (to the periphery) VNS on Days 1 and 2 (Efferent VNS group).

The experimental groups consisted of three following categories.

1) Control and LPS groups: Fifteen fetuses were used as controls receiving NaCl 0.9%. Twenty-three fetuses received LPS (100 n=2, 200 n=1, 400 n=15 or 800 n=5 as ng/fetus/day) derived from E. coli (Sigma L5293, from E coli O111:B4, readymade solution containing 1 mg/ml of LPS) were administered intravenously to fetuses on days 1 and 2 at 10.00 am to mimic high levels of endotoxin in fetal circulation over several days as it may occur in chorioamnionitis. As we identified that IL-6 response did not depend on the LPS dose in the applied range,(Durosier et al., 2015) these animals were all considered as one LPS group for statistical comparison's purposes.
2) Vx+LPS groups: Eleven animals were vagotomized (Vx) and exposed, similar to the LPS group, to LPS400 (n=6) or LPS800 (n=5).
3) Efferent VNS group: Eight additional Vx animals were subjected to bilateral cervical VNS applied via NeuroLog's NL512/NL800A using pulse sequence pre-programmed in Spike 2 for 10 minutes prior to and 10 min after each injection of LPS. The VNS settings were as follows: DC rectangular 5 V, 100 uA, 2 ms, 1 Hz according to (Borovikova et al., 2000). VENG was recorded at 20,000 Hz.(Castel et al., 2021b)

## Cardiovascular analysis

Systolic, diastolic and mean ABP (sABP, dABP and mABP) and mean FHR were calculated for each animal, at each time point (baseline, 1h, 3h, 6h, 24h, 48h and 54h), as an average of the artifact-free 30 preceding minutes (60 preceding minutes for the baseline) using Spike 2 (Version 7.13, CED, Cambridge, U.K.).

## Cytokine analyses

Cytokine concentrations (IL-6) in plasma were determined by using an ovine-specific sandwich ELISA. Mouse anti sheep monoclonal antibodies (capture antibody IL-6, MCA1659, Bio Rad AbD Serotec) were pre-coated at a concentration 4 µg/ml on ELISA plate at 4°C for overnight, after 3 times wash with washing buffer (0.05% Tween 20 in PBS, PBST), plates were then blocked for 1h with 1% BSA in PBST. Following 3 times washing, 50 µl of serial diluted protein standards and samples were loaded per well and incubated for 2 hours at room temperature. All standards and samples were run in duplicate.

Recombinant sheep proteins (IL-6, Protein Express Cat. no 968-305; TNF-α, Cat. no 968-105) were used as ELISA standard. Plates were then washed 3 times. Rabbit anti sheep polyclonal antibodies (detection antibody IL-6, AHP424, Bio Rad AbD Serotec) at a dilution of 1:250 were applied in wells and incubated for 30 min at room temperature. Plates were then washed with a washing buffer for 5 times. Detection was accomplished by assessing the conjugated enzyme activity (goat anti-rabbit IgG-HRP, dilution 1:5000, Jackson ImmunoResearch, Cat. No 111-035-144) via incubation with TMB substrate solution (BD OptEIA TMB substrate Reagent Set, BD Biosciences Cat. No 555214), color development reaction was stopped with 25 µl of 2N sulphuric acid. Plates were read on ELISA plate reader at 450 nm, with 570 nm wavelength correction (EnVision 2104 Multilabel Reader, Perkin Elmer). The sensitivity of IL-6 ELISA was 16 pg/ml. For all assays, the intra-assay and inter-assay coefficients of variance was <5%.

## Immunofluorescence analyses of the terminal ileum

Briefly, we quantified CD11c+ expression (i.e., M1 macrophages) normalized for CD11c+ cell counts as published.(Liu et al., 2016b)

Terminal ileum tissue samples of approximately 10 cm in length were taken from the fetus during necropsy and immediately immersed in 4% paraformaldehyde (PFA) for 48 to 72 hours. The tissue samples were then washed and stored in 1x phosphate-buffered saline (PBS) changed daily for 3 days. Finally, the samples were stored in 70% ethanol until further processing and kept at 4 °C when they were in liquid. After that they were processed with Leica TP 1020 Automatic Tissue Processor (Leica Instruments, Nussloch, Germany) and then embedded in paraffin with Leica EG 1160 Paraffin Embedding Center (Leica Microsystems, Nussloch, Germany). 5 micrometers slices were obtained from slicing the embedded tissue samples with the Leica RM2145 Rotary Microtome (Leica Microsystems, Nussloch, Germany), and mounted on the Fisherbrand Colorfrost Plus microscope slides (Fisher Scientific).

To detect M1 macrophages, we stained the ileum tissues with a mouse anti-sheep monoclonal antibody CD11c (1:50 dilution; RTI, LLC, Brookings, SD; Cat. no CD11C 17–196).

All slides were stained simultaneously with the same solutions to minimize staining variations. Slides were deparaffinized with three 5-minute washes in xylene, rehydrated for 2 minutes each through descending ethanol series (100%, 100%, 90%, 90% and 70%), then rinsed in deionized water for 5 minutes. Tissue sections were then subjected to antigen retrieval in 10 mM sodium citrate at pH 6.0 in a pressurized antigen retriever (Electron Microscopy Sciences, Hatfield, PA) followed by 3 phosphate-buffered saline (PBS) rinses of 5 minutes each, then all subsequent steps were performed in a closed humidity chamber. Non-specific protein binding was blocked with Background Sniper protein blocker (Biocare Medical, Pacheco, CA) for 10 minutes, then slides were incubated overnight at 4℃ with the respective primary antibody as outlined above. Following 3 further PBS washes, slides were incubated for 40 minutes at room temperature with Alexa 568 goat anti-mouse IgG (Thermo Fisher, Waltham, MA). After another 3 x 5-minute PBS rinse, a 2-minute counterstain with DAPI, and further PBS rinse, the slides were coverslipped with Prolong Gold mounting media (Thermo Fisher/Invitrogen, Waltham, MA).

Negative controls were performed by replacing the primary antibody step with purified pre-immune mouse IgG to rule out non-specific binding.

<u>Imaging and Analysis:</u> Quantification was performed on six randomly selected high-power fields (40x Oil magnification). Images were captured using a Zeiss AxioImager Z1 microscope (Carl Zeiss Canada, Toronto, Canada). Identical illumination settings were used for all images, and analysis was performed using Image Pro Premier 9.2 software (Media Cybernetics Inc., Rockville, MD), with the analyst blinded as to the animal group. After preliminary testing of a random sample of images (including screening against negative controls), binary intensity thresholds and settings were established to select and count only cells and cell fragments deemed to be positively stained. Semi-automated macros were then written in the analysis software to minimize bias and ensure consistency.

The combined pictures were used to quantify the CD11c+ macrophages as described below. The complete tissue was scanned, and six to seven individual pictures were cropped for quantification with Image Pro Premier.

We evaluated the terminal ileum tissues for the presence of CD11c+ cells in two ways. First we quantified the density of the intensity of the CD11c+ stains (area of the stains divided by the total area of the individual HPF, “range mean”) with the Image Pro Premier 9.2 software (version 9.2; Media Cybernetics, Rockville, MD).

Second, we counted the number of CD11c+ cells as follows. After preliminary testing of a random sample of images (including screening against negative controls), binary intensity thresholds and settings were established to select and count only cells and cell fragments deemed to be positively stained. Semi-automated macros were then written in the analysis software to minimize bias and ensure consistency.

Since fetal gut tissue is full of erythrocytes and other autofluorescent components, a strategy to eliminate counting of any non-specific fluorescence was developed. As autofluorescence appears in both the green and red channels, while our true CD11c signals only appear in the Alexa 568/red channel, we used the green channel to create a binary “exclusion” mask. This was achieved as follows: greyscale versions of the green channel were subjected to automated, binary thresholding using the minimum variance “auto bright” algorithm in Image Pro. A count was performed to create outlines of those counted spots (mostly erythrocytes). The counted spots were grown by 7 pixels to encompass any convolved signal immediately surrounding them. The outlines of these counted zones were then geographically superimposed onto the corresponding multichannel image. Following this, a Boolean exclusion was imposed, so that subsequent counting of the red CD11c signal would only take place in the regions of interest outside of the autofluorescent cellular area.

After imposing this region of interest selection, the multichannel image was slightly sharpened using a High Gaussian filter to yield better morphological data. To select and measure the CD11c positive cells, a binary threshold was created using the “smart selection” tool in Image Pro, which selectively recognized the red cells based on color, morphology, and tissue background characteristics. CD11c positive cellular debris seemed to be evident in several animals, so we chose to analyze cells and debris or “spots” separately. Cells were considered to be anything above 350 pixels in size. Data from the debris was not included in morphological statistics. All cell and spot measurements were normalized to the tissue area. To determine this area, a manually created binary intensity threshold was used to identify and measure the dark background surrounding tissue, and this area was subsequently subtracted from the total image area to yield a measurement of the total tissue area in the image.

Finally, for each animal, the density of intensity of CD11c+ cells per HPF determined in the first step above was normalized for the number of cells per HPF computed in the second step by calculating a ratio of the two. *The latter ratio was used for the statistical analyses, reported for simplicity as CD11c+ cells.*

## Insulin analyses

Insulin concentrations in fetal sheep plasma were determined by using an ovine-specific insulin ELISA kit (ALPCO Diagnostics, Cat. No. 80-INSOV-E01, Salem, NH). Briefly, fetal sheep blood samples were collected with a heparinized syringe at baseline, 3h, and 48h; plasma samples were obtained by centrifugation at 4000g, 4°C for 4 min and stored in aliquots at -80°C until assay. Twenty-five (25) microliters plasma were loaded per well without dilution in duplicates. The sensitivity of the assay was 0.14 ng/ml; the intra-assay and inter-assay coefficients of variance were <5.96% and <5.78%, respectively.

## Statistical analysis

General linear modeling (GLM) in Exploratory/R was used to assess the effects of treatment (LPS, Vx, Vx+efferent VNS+LPS) while accounting for repeated measurements of fetal blood gasses, glucose, lactate, acid-base status, insulin, fetal cardiovascular responses (blood pressure and heart rate), fetal systemic inflammatory response (plasma IL-6 cytokine), and fetal terminal ileum inflammatory response (CD11c+ cells). Consequently, experimental groups and time points served as predictor variables; as base levels served the control group and the baseline time point, respectively. Coefficients of estimates were used to rank the statistically significant differences by their magnitude (i.e., treatment effect larger or smaller) and direction (i.e., increase or decrease). Not all measurements were obtained in each animal. In such a case, the sample size is reported explicitly.

All results are presented as median±IQR (interquartile ranges) with a significant difference assumed for $P<0.05$. Statistically significant correlations are reported for Spearman $R2>0.1$.

# Results

### Blood gas and metabolites

This experimental fetal sheep cohort's morphometric, arterial blood gasses, acid-base status, cardiovascular characteristics and cytokine responses have been reported in part: for control and LPS groups as well as for acute post-Vx effects.(Castel et al., 2021a; Durosier et al., 2015) Briefly, fetal arterial blood gasses, pH, BE and lactate were within physiological range during the baseline in both groups.

We report here, for the first time, the findings in the Vx and efferent VNS groups over the 54h course of low-dose LPS-triggered fetal inflammatory response (Table 1). The respective baseline measurements in the groups of vagotomized animals did not differ. We report changes in glucose in a dedicated subsection below. In Table 1 we show some detectable, yet physiologically not meaningful treatment group and time effects for pH, $pO_2$, $pCO_2$, $O_2Sat$, and BE (p values ranging from <0.001 and 0.3; statistically significant changes are identified in Table 1). Of note, there was a ~40% elevation of lactate in all LPS-treated groups regardless of vagus nerve manipulation at 3 and 6 hours post-LPS corresponding to the peak of fetal inflammatory response indicating a mild metabolic acidemia.

### Cardiovascular responses

Overall, we observed no overt cardiovascular decompensation due to LPS-triggered inflammatory response (Table 2). There was a consistent 6 hours time effect for FHR and ABP corresponding to the acute fetal inflammatory response after the first dose of LPS, regardless of the treatment group (LPS without or with vagus nerve manipulation).

Specifically, for FHR, there were no significant treatment effects, but there was a mild heart rate increase without overt tachycardia observed as a time effect for 6 ($p<0.001$) and 48 hours ($p=0.03$), corresponding to the immediate 6 hours post first dose of LPS administration and 24 hours after the second dose. The 6 hours effect was twice more pronounced than 24 hours.

For dBP, there was a mild decrease due to LPS treatment alone ($p=0.01$), but not due to vagus nerve manipulation. There was also a similar mild decrease as a time effect at 6 and 30 hours ($p=0.003$ and $p=0.046$, respectively), i.e., 6 hours post the first and second LPS doses.

For mBP, there was only a mild decrease as a time effect at 6 hours regardless of the treatment ($p=0.009$).

For sBP, we observed again a time effect-related mild decrease at 6 hours ($p=0.016$) and treatment-related effects with mild increases for Vx+LPS800 ($p=0.02$) and the efferent VNS groups ($p=0.046$).

### Systemic inflammatory response: IL-6

LPS provoked a systemic inflammatory response with time effect, i.e., peaking, at 3 and 6 hours (both p values <0.001), as measured by fetal arterial IL-6 concentrations over 3 days (Fig. 2). Surprisingly, Vx+LPS400 restored the levels of IL-6 to control levels, while treatment effects with elevated IL-6 were observed for Vx+LPS800 ($p<0.001$), LPS ($p<0.001$) and efferent VNS ($p=0.003$) groups, in the decreasing order of magnitude.

Of note, in the efferent VNS group, the inflammatory response abated quicker than in other LPS-exposed groups, while in the Vx+LPS800 group the response persisted the longest, up until 54 hours post LPS, the last experimental measurement time point.

Last but not least, we observed a clearly diametrical temporal profile of the fetal inflammatory response to LPS in Vx+LPS400 versus Vx+LPS800 group suggesting a sigmoid functional relationship between the vagal cervical denervation and the LPS dose–dependent magnitude of the fetal inflammatory response.

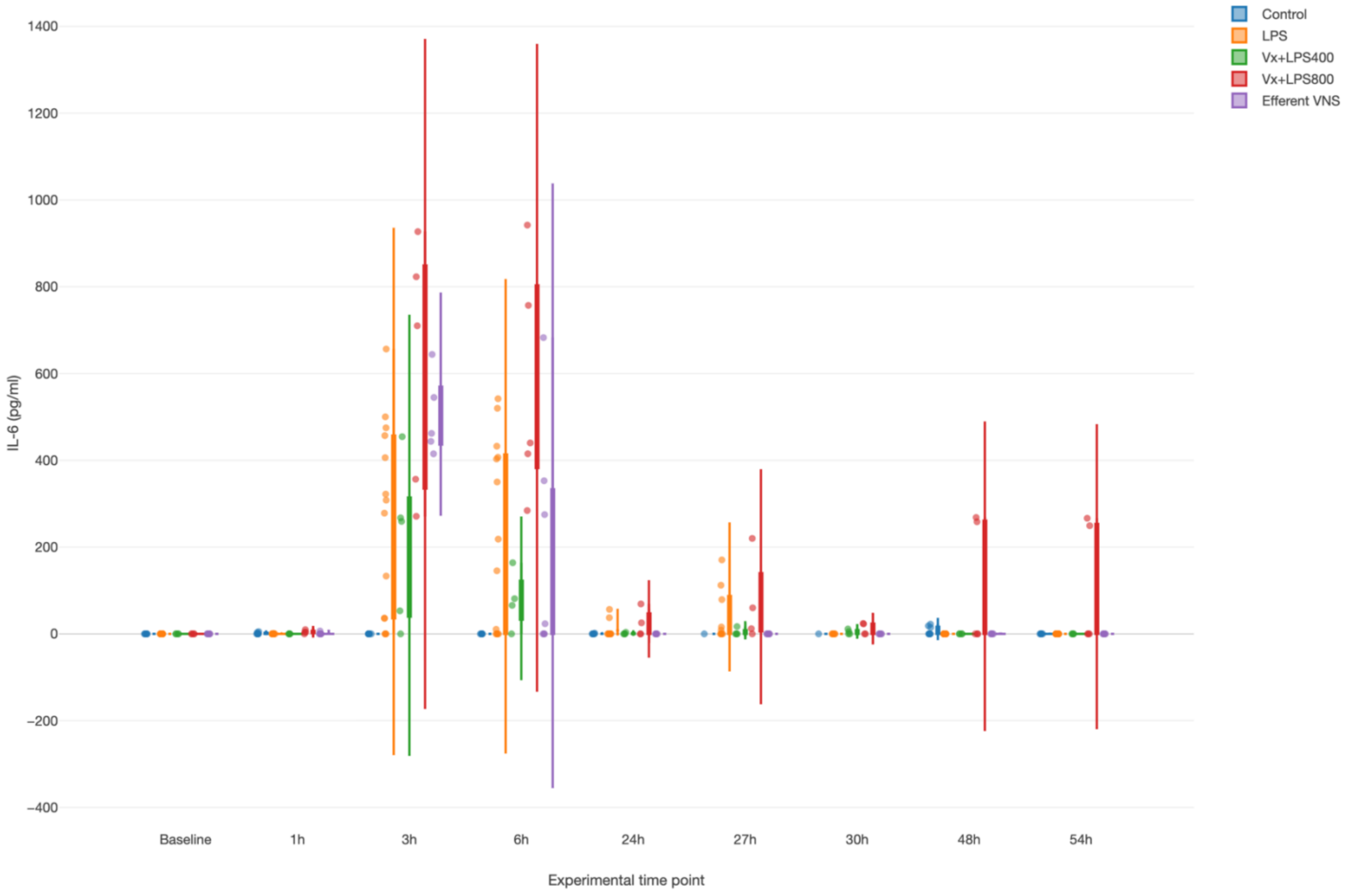

**Figure 2. Fetal systemic inflammatory response to intravenous LPS injection after baseline and at 24 hours: the impact of vagus nerve manipulation.**

Vx, bilateral cervical vagotomy during surgical instrumentation; LPS400 and LPS800 indicate the respective intravenous dose of LPS in ng/fetus/day given after baseline and 24 hours later; Efferent VNS group received Vx and LPS400 as the Vx+LPS400 group followed by VNS treatment around the LPS administration at days 1 and 2.

**Effects of vagus nerve manipulation on fetal systemic arterial glucose and insulin levels**

For glucose, we found a treatment effect for both Vx+LPS400 and Vx+LPS800 (Table 1). Since there was no significant difference between Vx+LPS400 and Vx+LPS800 insulin or glucose values at the respective time points, we combined them into one Vx+LPS group for further analysis in this subsection: there was only one positive pronounced effect for treatment in the Vx+LPS group ($p<0.001$) corresponding to a ~30% rise in glucose values throughout the experiment in this group (Fig. 3).

For insulin, we found positive and similar effects for time at 3h ($p<0.001$) and for treatment in the Vx+LPS group ($p=0.009$) and the efferent VNS group ($p=0.015$). In the Vx+LPS group, the rise was ~2.3-fold at 3 hours and ~1.3-fold rise in the efferent VNS group.

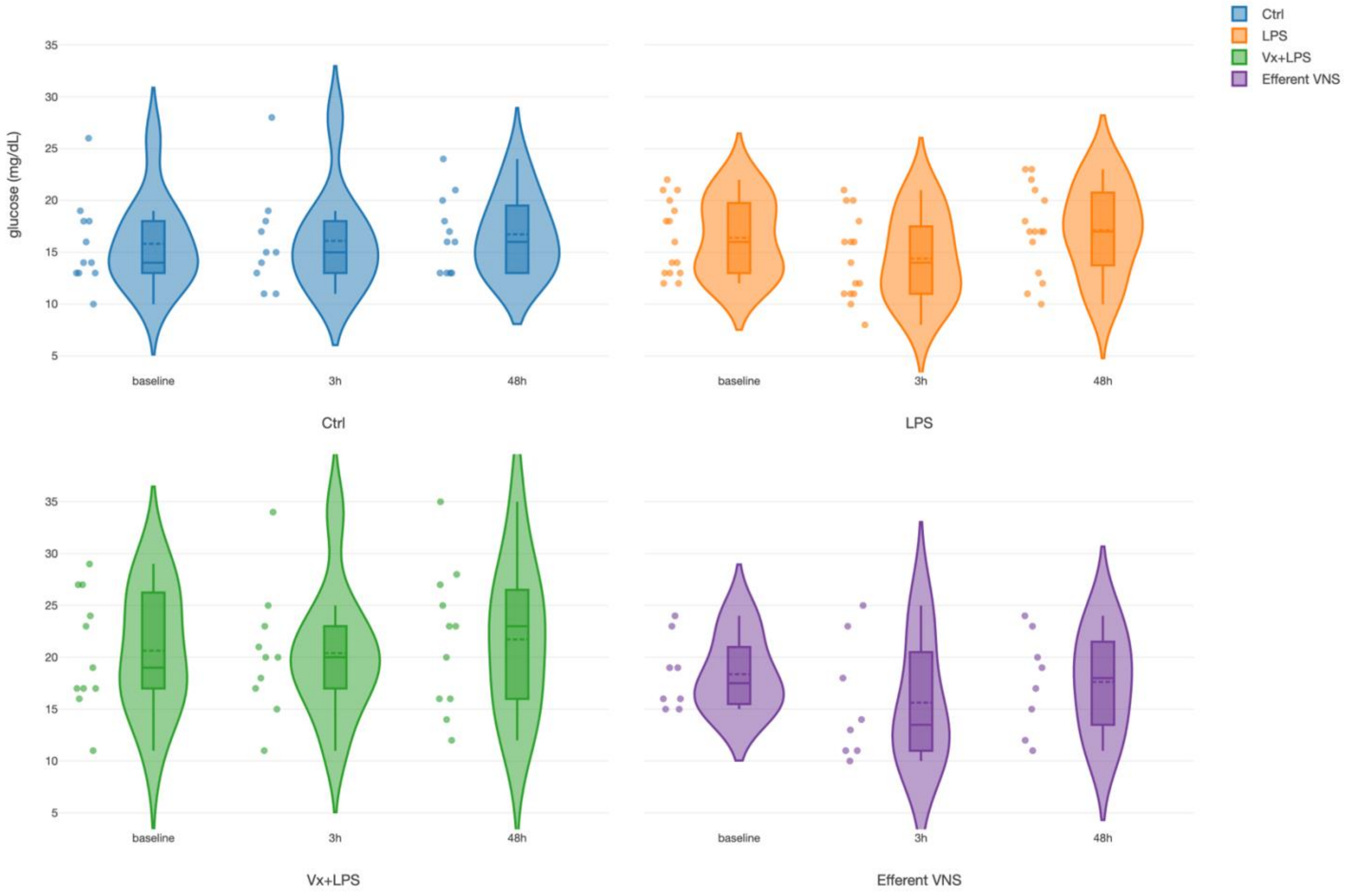

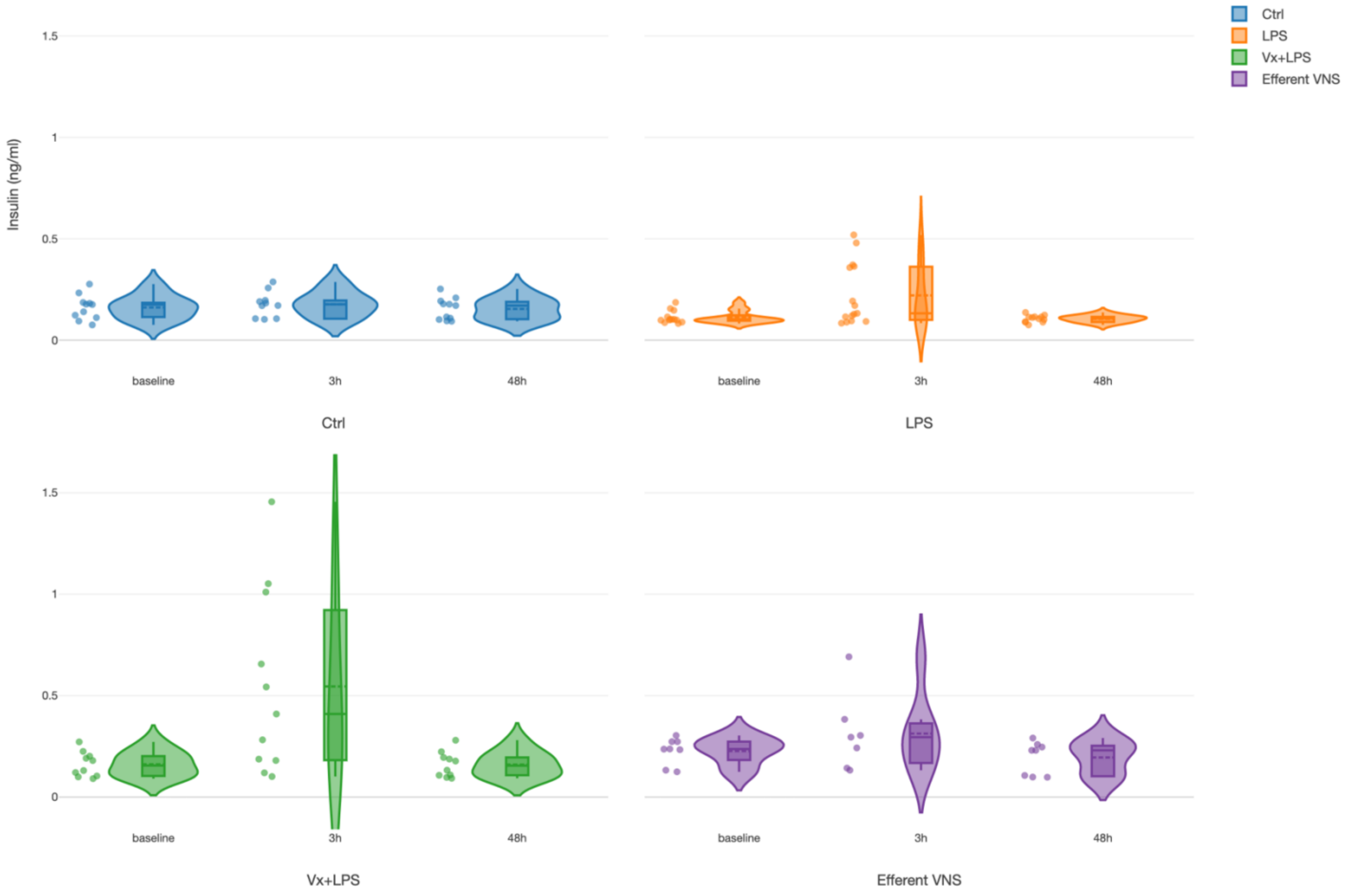

**Figure 3. Fetal systemic arterial glucose (TOP) and insulin (BOTTOM) dynamics** during fetal systemic inflammatory response to intravenous LPS injection: the impact of vagus nerve manipulation. Key time points are shown: baseline (prior to LPS injection), 3 hours and 48 hours post first LPS injection.

Ctrl, Control (no LPS, no Vx, vagotomy); Vx+LPS, bilateral cervical vagotomy during surgical instrumentation, the groups Vx+LPS400 and Vx+LPS800 were combined here due to lack of differences in glucose and insulin behaviors with either LPS dose; the LPS of 400 or 800 indicates the respective intravenous dose of LPS in ng/fetus/day given after baseline and 24 hours later; Efferent VNS group received Vx and LPS400 as the Vx+LPS400 group followed by VNS treatment around the LPS administration at days 1 and 2.

The above findings show that intact vagus nerve's activity modulates glucose homeostasis under conditions of LPS-induced inflammatory response. To test this further and explore whether insulin rise had direct anti-inflammatory effects we performed group-wise correlation analyses of glucose, insulin and IL-6 combining 3h and 48h time points, i.e., time points when these biomarkers reflected the evolving inflammatory response.

First, glucose correlated positively to insulin in all experimental groups except for LPS and Vx+LPS800 groups. Interestingly, the highest values of correlation were observed in the control, followed by the efferent VNS groups (Spearman R2=0.47 and 0.29, respectively). Vx+LPS400 group showed Spearman R2=0.22.

Second, we observed no correlations between IL-6 and insulin in any experimental group.

Together, the findings suggest a feed-forward relationship between insulin and glucose that is disrupted and reduced by vagal denervation, but no direct association between insulin and IL-6. Under conditions of Vx, LPS-triggered fetal systemic inflammatory and glycemic responses act synergistically, in part, in LPS-dose-dependent manner.

### Effects of fetal vagus nerve manipulation on terminal ileum's inflammatory response to LPS

CD11c+ cells (M1 macrophages') were increased in the LPS group (main effect LPS, p=0.003), as demonstrated in an earlier report,(Liu et al., 2016b) as well as in Vx+LPS800 (p=0.001), but not in the Vx+LPS400 or efferent VNS groups (Fig. 4).

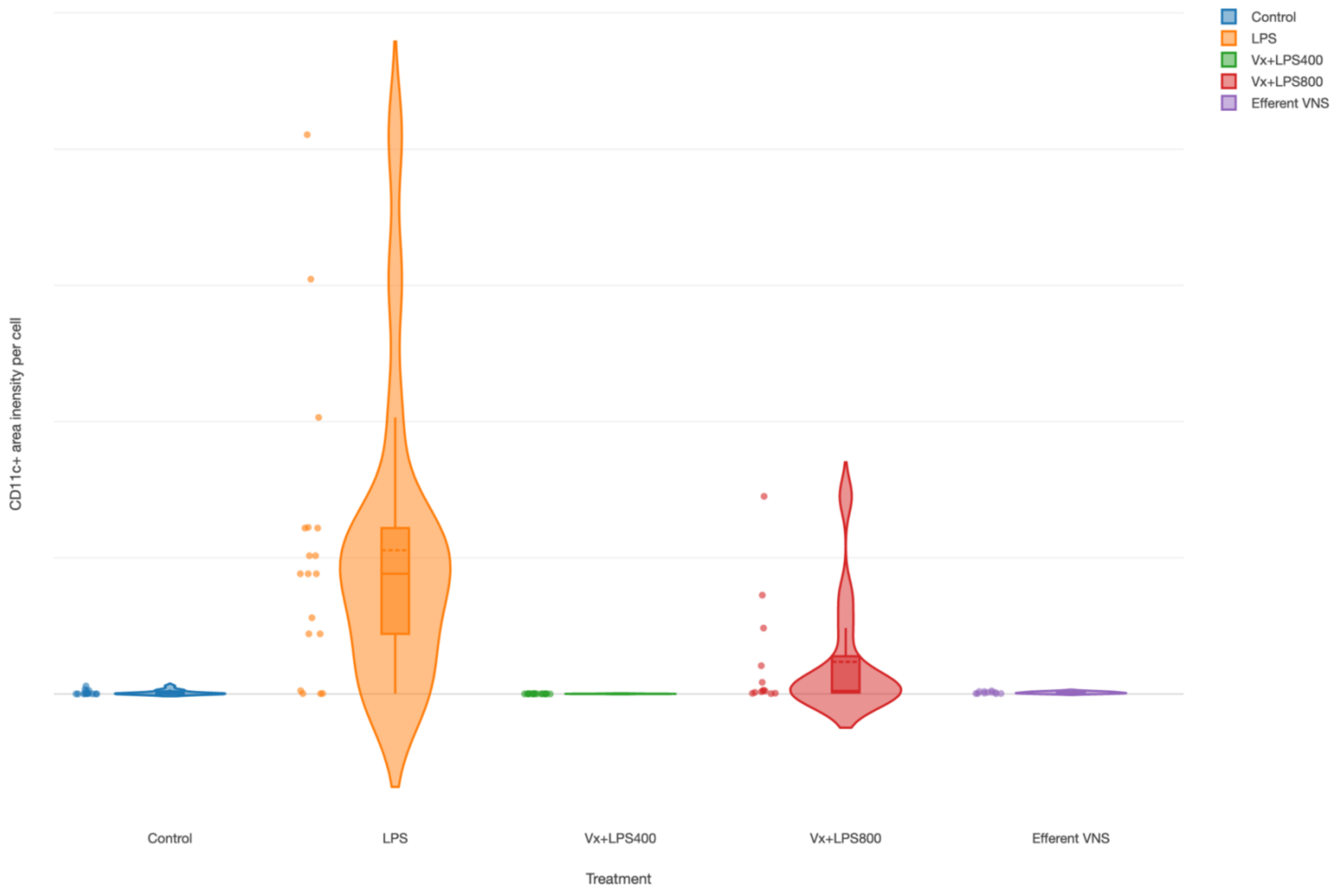

**Figure 4. Fetal terminal ileum inflammatory response to LPS and vagus nerve manipulation**: impact of vagal denervation (vagotomy) and efferent vagus nerve stimulation (VNS) on M1 (CD11c+ cells) macrophage behavior. **CD11c total area normalized by cell count** (pixels). GLM: Increase; main effect - LPS and Vx+LPS800. Control (n=5), LPS (n=12), Vx+LPS400 (n=6), Vx+LPS800 (n=4), efferent VNS (n=5).

## Discussion

We present findings on the effects of a mechanical definitive abrogation of the vagal signaling, the key neural substrate of brain-body communication, including the interruption of the cholinergic anti-inflammatory pathway, via the bilateral cervical vagotomy. This disruption alters the fetal systemic and gut regional immunometabolic responses to endotoxin exposure. The findings are summarized in Table 3 to facilitate the systematic interpretation.

### *General expected and novel effects of vagus nerve manipulation*

Despite some mild changes, our experimental cohort's morphometric, arterial blood gasses, acid-base status, and cardiovascular characteristics were within physiological range and representative for a late-gestation fetal sheep as a model of human fetal development near term.(Frasch et al., 2007; Rurak and Bessette, 2013) The effect of the low LPS dose we administered on the arterial blood gasses, acid-base status, and cardiovascular responses is compatible with mild septicemia (mild compensated metabolic acidemia and hypoxia) evidenced by a transient rise of IL-6 at 3 hours without overt shock with cardiovascular decompensation. The organ-specific effect of this inflammatory insult is evidenced by the rise of M1 type macrophages in terminal ileum. These systemic and organ-specific findings are in line with what is expected from fetal inflammatory response to endotoxin challenge and the efferent VNS diminishes these effects, again, as would be expected per our understanding of the function of the fetal cholinergic anti-inflammatory pathway.(Frasch et al., 2016; Garzoni et al., 2013)

In fetal sheep near-term, despite the high degree of maturity of the autonomic nervous system, Vx does not cause a compensatory rise in the activity of the sympathetic nervous system.(Herry et al., 2019) This was also evident in the present study with no rise in FHR or dABP measured. As such, we do not attribute the observed rise in glucose levels to the effects of the sympathetically mediated stress pathway.

Beyond these expected effects of vagus nerve manipulation, the present study confirms our glycemic control hypothesis: a complete withdrawal of peripheral vagus innervation results in 72 h delayed onset of sustained increase in systemic arterial glucose levels that begins *prior* to endotoxin exposure and persists unchanged for at least 54 h. Notably, this behavior is accompanied by a transient hyperinsulinemia at the peak of the systemic inflammatory response, and, unexpectedly, LPS-dose-dependent biphasic lower or higher levels of systemic and gut inflammation. An intermittent efferent bilateral cervical VNS restores these changes.

In summary, this study reveals three novel facets of the vagal control of inflammation and metabolism in near-term, physiologically mature, fetus: 1) the short-term immunometabolic regulatory role, 2) the hormetic behavior and, concurrently, 3) adaptive, longer-term, predictive computation manifesting as immunometabolic memory of the former exposure to endotoxin.

In the following, we discuss these observations and propose future experiments.

### *Immunometabolic effects of vagus nerve manipulation*

Vx caused chronic rise in glucose levels, notably, starting prior to and regardless of LPS while efferent VNS restored this state to control levels. Meanwhile, Vx also caused transient hyperinsulinemia at the peak of LPS-triggered IL-6 rise, but not at baseline; efferent VNS restored this in part. The group-wise positive glucose-insulin correlations decreased from control to efferent VNS to Vx+LPS400 (but not LPS or Vx+LPS800) groups. Together, these findings demonstrate that Vx disrupts glucose-insulin homeostasis. The fact that the efferent VNS restored the hyperinsulinemia only partially indicates the synergistic effect of inflammation on the disruption of the glucose-insulin homeostasis when vagal control is absent or diminished. Despite these observations, we did not observe any clear correlations between glucose levels and M1 macrophage activity in the terminal ileum or between insulin and IL-6 or insulin and M1 macrophage activity in any experimental group. This would indicate that the triggered rise in glucose levels was not sufficient to cause injury. Alternatively, the mechanisms of hyperglycemic injury and the anti-inflammatory effects of insulin involve causal pathways not captured in the present study.

Anorexic or food intake increasing signals are those that predict future energy requirements.(Myers et al., 2021) The emphasis on “prediction” is of great importance as it recapitulates the notions of predictive brain put forward by Peters *et al.* in other systems regulatory contexts.(Peters et al., 2017) Signals that predict increased fuel demand, e.g., cold exposure, increase both hunger and energy expenditure. This line of thought warrants further development in the context of neurally integrated immunometabolic control.

In such a framework, does inflammation signal metabolic demand which is controlled, under physiological conditions, by insulin via the vagus nerve, obscuring the net increase in insulin secretion? Under the disrupted conditions of Vx, inflammation would trigger gluconeogenesis, increase in systemic glucose availability, as we see here, due to lacking or diminished central feedback on the beta cells/liver’s insulin production via the vagus nerve, we detect hyperinsulinemia. This is supported by literature reporting that VNS triggers insulin production and reduction of glucose levels.(Frohman et al., 1967; Lee and Miller, 1985) Afferent VNS fibers, activated by non-selective VNS, are thought to produce chronic hyperglycemic effects, while VNS of strictly efferent fibers mediates hypoglycemic effects.(Meyers et al., 2016; Stauss et al., 2018) In the present study, strictly efferent VNS was performed and the findings align with the literature.

Reviewing the evidence of the past ~15 years, Meyers *et al.* proposed most recently that brain systems that control various peripheral metabolic systems, such as the activities of the adipose tissues and glucose control systems, should be viewed, together, as the components of a larger, highly integrated, 'fuel homeostasis' control system.(Myers et al., 2021) This view places the vagus nerve with its ubiquitous distributed afferent sensing and efferent fibers at the center of this fuel homeostasis control system. The authors note that the developmental aspect of adult metabolic phenotype requires further elucidation as there is little doubt that it plays a key role in the puzzle of metabolic function in health and disease throughout the lifespan.

There is further a recognition that immune and metabolic regulatory systems are in fact two sides of the same coin, an integrated, adaptive, multi-scale, molecular and systems-level network.(Castel et al., 2021a; Frasch et al., 2018, 2019, 2020) This understanding has led to characterizing peripheral macrophages and, increasingly, also the central nervous system’s glia by their immunometabolic phenotype in response to various stressors.(Cao et al., 2015, 2019; Cortes et al., 2017; Desplats et al., 2019; Lajqi et al., 2021; Lang et al., 2020; Lee et al., 2018)

Specifically for this developmental stage and its animal model, we proposed a scale-invariant concept that modeling the integrated system’s level responses to manipulation of vagus nerve signaling under conditions of repeated endotoxin exposure is akin to modeling the behavior of glial cells in vitro in response to manipulation of α7 nicotinic acetylcholine receptor (α7nAChR) under conditions of single or multi-hit LPS exposure (Fig. 5).(Castel et al., 2021a; Frasch et al., 2018) In such a framework, the response pattern symmetry (or invariance) across scales of organization, from cellular to system’s, appears to be evident with regard to the changes in immunometabolic behavior. From the teleological evolutionary standpoint, such a phenomenon appears to provide adaptive benefits to complex organisms at all scales of organization by optimizing the available and predicted energy utilization and entropy production rates via immunoceptive inference which follows the free energy principle.(Bhat et al., 2021; Friston, 2010; Yang et al., 2021)

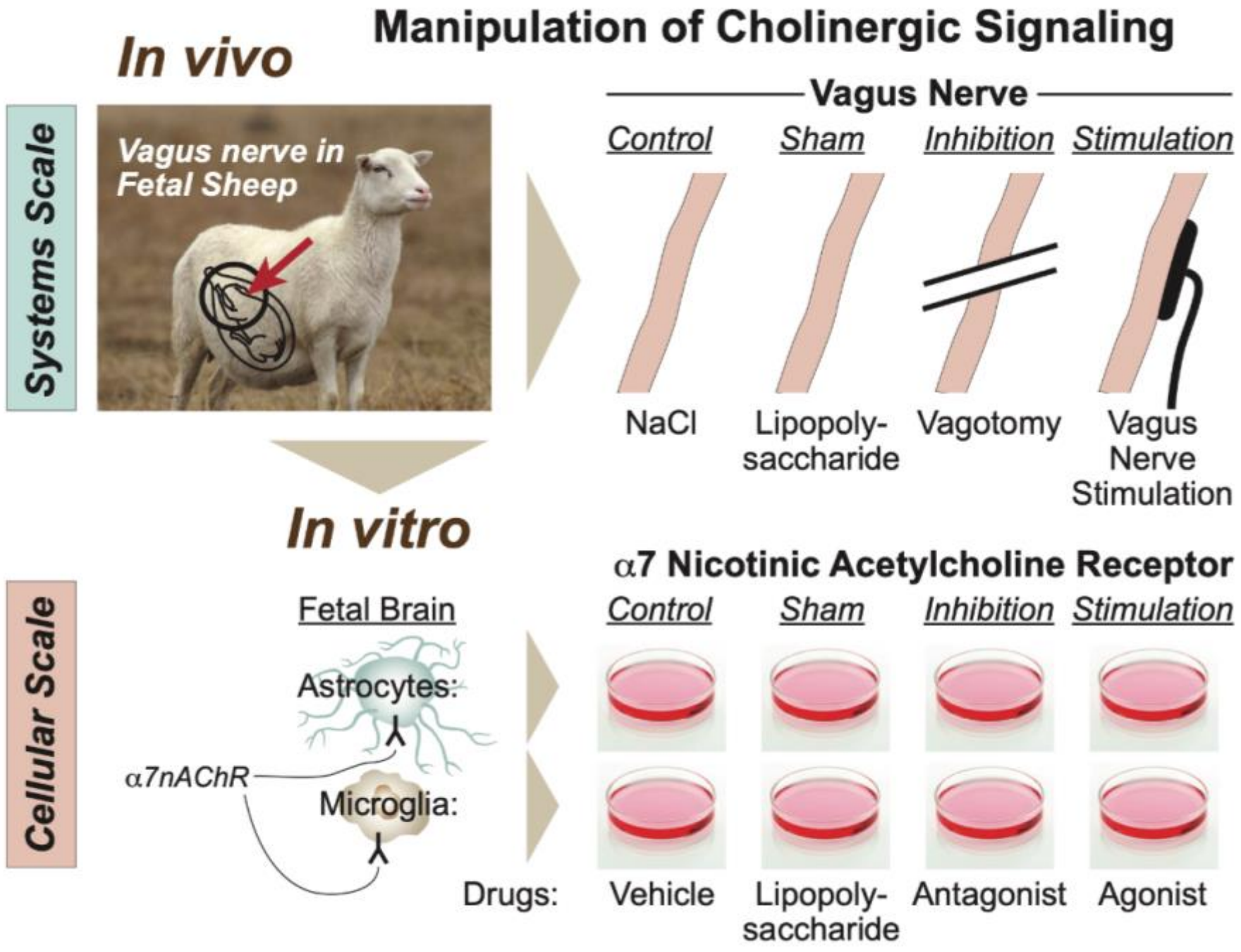

**Figure 5. Modeling physiological scales of organization of cholinergic signaling.** (Adopted from (Frasch et al., 2018) with permission).

As a methodological limitation, we point out we measured insulin at three time points only. This limits the conclusions on the temporal dynamic relationship between insulin and other biomarkers and requires further studies with higher temporal resolution of insulin in relation to glucose and IL-6, in particular, in response to an inflammatory stimulus such as LPS. Given the potential of insulin to reduce inflammation, a deeper understanding of the physiology and pathophysiology of this dynamic relationship is of great translational interest.

More long-term, there is a well-established need for preventive early-life interventions to stem the epidemics of obesity.(Levin, 2006b) Our findings suggest that VNS may emerge as one such early, postnatal intervention to reset the "metabolic imprinting": the effect of VNS showed delayed at 48 (but not yet at 3) hours. That suggests that chronically there may be a resetting influence of VNS as treatment. This needs to be explored further.

### *Hormesis*

We reported the increase in M1 macrophage activity in the terminal ileum in the LPS group.(Liu et al., 2016b) Here we demonstrate that a complete withdrawal of vagal innervation followed by LPS exposure restores M1 activity to control levels. Efferent VNS had a similar anti-inflammatory effect. Of note, we observed a consistent LPS-dose-dependent hormetic behavior under Vx for both the systemic inflammatory response pattern and the regional terminal ileum inflammatory response.

The hormetic behavior is represented by a functional dependency of anti- or pro-inflammatory response to endotoxin depending on its dose. Vx followed by LPS400, but not by LPS800 regimen reduced systemic and gut-specific inflammation levels. That endotoxin dose-response pattern is compatible with hormetic behavior of complex systems and is thought to reflect an adaptive or computationally predictive response observed widely in biological systems.(Calabrese and Mattson, 2017) This is the first demonstration of such behavior under conditions of an ablated cholinergic anti-inflammatory pathway.

Remarkably, the immunometabolic memory of endotoxin exposure is reflected in the hormetic behavior. The surprisingly anti-inflammatory effect of Vx+LPS400 is even more pronounced at the second endotoxin exposure (at 24 h of the experiment) and contrasted by the absence of any habituation for Vx+LPS800 group. This is accompanied by higher levels of regional inflammation in the terminal ileum in this group. The efferent VNS appears to boost the memory effect for the Vx+LPS400 group. As this finding was surprising and animal numbers were already planned for, we were unable to test this by adding Vx+LPS800+efferent VNS treatment as done for Vx+LPS400 group. As such, this remains an important follow-up study to be made: does efferent VNS in Vx+LPS800 fetuses restore the memory effect boosting habituation to the second endotoxin exposure? To further explore the discovered hormesis systematic studies are needed on what the endotoxin dose-response relationship is depending on the maturity and species type.

### *Implications for NEC etiology and avenues of treatment*

Impaired mesenteric oxygenation due to poor perfusion or cyanosis are considered the major predisposing factors for NEC. While the etiology of NEC in term infants remains unknown, the contribution of chorioamnionitis, often subclinical, is discussed as one of the contributing factors. A common event is the exaggerated inflammatory response to an exogenous trigger as a prequel to NEC. This leads to increased intestinal permeability and mucosal injury. (Caplan and Hsueh, 1990; Lin and Stoll, 2006) Higher levels of cytokines such as IL-6 correlate with the severity of NEC in preterm neonates and contribute to chemotaxis.(Edelson et al., 1999; Harris et al., 1994; Sharma et al., 2007)

Both short-term (hours) and chronic hyperglycemia and hypoinsulinemia promote organ injury and inflammation.(Jafar et al., 2016; Zhu et al., 2018) Our findings show that this relationship depends on the intact vagus nerve signaling. The acute, transient rise of insulin in the vagotomized animals at 3h post LPS was not fully recovered by efferent VNS indicating the synergistic role of inflammation and hyperglycemia in this response.

Counterintuitively, it appears plausible that lowering the cholinergic tone in conditions of neonatal hyperglycemia and/or mild, but not severe, septicemia can help prevent NEC or mitigate its deterioration. Future studies should explore the predictive potential of monitoring and manipulating the vagal activity in neonates, perhaps via the heart rate variability (HRV) monitoring or via direct non-invasive vagus electroneurogram.

### *Significance and perspectives*

The key translational implications of our findings for neonatal healthcare are the potential of selective efferent VNS to reduce acute systemic and regional gut inflammation as well as the metabolic effects of the efferent VNS. The latter appear to have complex effects on several time scales. Acutely, the metabolic effects may aid reduce inflammation boosting the vagally mediated immunometabolic regulatory networks. Chronically, the effects may be in resetting in-utero programming of susceptibility to obesity. The latter aspects were not the subject of the present study but appear to be a logical speculative extension of the findings and require further dedicated experiments.

The surprising hormetic, LPS-dose-dependent, systemic and ileum anti-/pro-inflammatory effect of Vx warrants further studies. Whether Vx alone or its synergy with the LPS-induced inflammatory response are responsible for the observed metabolic behaviors also requires further studies. The intermittent rise of insulin concomitant with the rise of IL-6 suggests a synergistic immunometabolic pattern.

Our results are relevant for a broader understanding of immunometabolic programming in the perinatal stage of development and the possibility of using VNS to reset it. Implications for understanding the involvement of the vagus nerve in glucosensing and inflammation warrant further investigation in general and regarding therapeutic opportunities for NEC in particular.

## Additional information

### Competing Interests

None of the authors has any conflicts of interests to declare.

### Funding

Funded by CIHR, FRQS, Molly Towell Perinatal Research Foundation (to MGF).

### Acknowledgements

The authors gratefully acknowledge technical support by Esther Simard, Marco Bosa, Pierre-Yves Mulon and L. Daniel Durosier during the experiments. We thank Dr. Andrew Seely for fruitful conversations about the findings and the idea of the dose dependent biphasic response pattern under vagotomy and endotoxin exposure. We thank Tziporah Thompson for the masterful execution of the visual abstract.

# Tables

**Table 1. Arterial blood gases and metabolites** during LPS-induced fetal inflammatory response and vagus nerve manipulation (Median and interquartile range, IQR)**.**

| | | Experimental group | | | | | | | | | |
|---|---|---|---|---|---|---|---|---|---|---|---|
| **Parameter** | **Time point** | **Control** | | **LPS** | | **Vx_LPS_400** | | **Vx_LPS_800** | | **Efferent VNS** | |
| **pH** | Baseline | 7.37 | *0.05* | 7.37 | *0.05* | 7.40 | *0.07* | 7.40 | *0.01* | 7.38 | *0.03* |
| | 1h | 7.37 | *0.04* | 7.36 | *0.07* | 7.39 | *0.06* | 7.38 | *0.04* | 7.37 | *0.05* |
| | **3h** | 7.36 | *0.06* | 7.34 | *0.09* | 7.36 | *0.08* | 7.36 | *0.01* | 7.35 | *0.02* |
| | **6h** | 7.37 | *0.04* | 7.35 | *0.08* | 7.35 | *0.07* | 7.35 | *0.03* | 7.36 | *0.07* |
| | 24h | 7.36 | *0.05* | 7.36 | *0.04* | 7.39 | *0.06* | 7.40 | *0.03* | 7.37 | *0.03* |
| | 27h | 7.33 | *0.08* | 7.35 | *0.03* | 7.36 | *0.05* | 7.39 | *0.03* | 7.37 | *0.04* |
| | 30h | 7.35 | *0.03* | 7.35 | *0.03* | 7.35 | *0.06* | 7.40 | *0.03* | 7.37 | *0.05* |
| | **48h** | 7.34 | *0.07* | 7.35 | *0.07* | 7.37 | *0.03* | 7.38 | *0.02* | 7.37 | *0.03* |
| | **54h** | 7.34 | *0.07* | 7.35 | *0.04* | 7.36 | *0.06* | 7.37 | *0.02* | 7.36 | *0.04* |
| | Median *IQR* | 7.36 | *0.05* | **7.35** | ***0.06*** | 7.37 | *0.06* | **7.38** | ***0.03*** | 7.36 | *0.04* |
| **paO2** | Baseline | 15.8 | *13.0* | 19.6 | *10.6* | 20.2 | *5.7* | 21.3 | *6.5* | 19.7 | *9.6* |
| | 1h | 17.2 | *16.2* | 19.0 | *8.5* | 20.6 | *6.0* | 24.0 | *1.4* | 18.6 | *7.8* |
| | 3h | 18.0 | *18.3* | 19.5 | *8.6* | 19.5 | *6.3* | 22.1 | *2.7* | 18.5 | *8.1* |
| | 6h | 18.4 | *19.5* | 17.3 | *8.5* | 17.7 | *9.3* | 19.4 | *4.6* | 16.9 | *8.9* |
| | 24h | 18.8 | *14.1* | 18.8 | *9.8* | 19.4 | *6.5* | 21.3 | *0.7* | 18.7 | *10.0* |
| | 27h | 18.2 | *7.0* | 20.4 | *7.2* | 20.9 | *9.8* | 22.9 | *2.4* | 17.9 | *7.0* |
| | 30h | 18.0 | *5.5* | 18.8 | *7.5* | 19.4 | *9.3* | 20.9 | *2.6* | 18.3 | *12.1* |
| | 48h | 17.9 | *11.1* | 18.2 | *9.3* | 20.7 | *8.9* | 18.7 | *11.6* | 19.2 | *8.9* |
| | 54h | 18.8 | *12.2* | 21.2 | *9.5* | 21.3 | *6.0* | 19.3 | *11.9* | 18.0 | *8.3* |
| | Median *IQR* | 17.9 | *13.0* | **19.2** | ***8.8*** | 20.0 | *7.5* | 21.1 | *4.9* | 18.4 | *9.0* |
| **paCO2** | Baseline | 51.2 | *6.2* | 51.6 | *6.9* | 48.6 | *8.4* | 49.4 | *3.3* | 49.9 | *2.4* |
| | 1h | 52.0 | *11.0* | 51.5 | *11.3* | 49.0 | *6.2* | 49.6 | *4.5* | 50.3 | *4.9* |
| | 3h | 51.3 | *10.2* | 51.6 | *11.3* | 51.0 | *8.9* | 52.0 | *2.3* | 51.4 | *4.6* |
| | **6h** | 52.8 | *12.1* | 53.1 | *8.5* | 53.0 | *9.9* | 54.7 | *0.8* | 52.6 | *6.3* |
| | 24h | 52.9 | *11.6* | 51.9 | *7.2* | 50.4 | *6.5* | 51.0 | *2.3* | 51.6 | *5.4* |
| | 27h | 56.5 | *1.1* | 50.7 | *5.8* | 50.0 | *11.3* | 49.5 | *2.7* | 52.7 | *5.5* |
| | 30h | 56.3 | *1.1* | 51.3 | *5.0* | 53.0 | *7.1* | 49.6 | *6.0* | 52.4 | *4.5* |
| | 48h | 52.8 | *6.0* | 52.0 | *6.1* | 49.7 | *7.1* | 50.5 | *5.5* | 52.1 | *4.9* |
| | **54h** | 54.5 | *11.9* | 50.5 | *10.5* | 52.6 | *6.3* | 51.9 | *3.9* | 52.6 | *5.3* |
| | Median *IQR* | 53.3 | *7.9* | 51.6 | *8.1* | 50.8 | *8.0* | **50.9** | ***3.5*** | 51.8 | *4.9* |
| **O2Sat** | Baseline | 44.8 | *32.7* | 52.5 | *29.3* | 50.4 | *21.9* | 60.4 | *10.2* | 54.6 | *35.5* |
| | 1h | 48.1 | *44.1* | 47.9 | *23.7* | 50.7 | *26.0* | 64.8 | *9.2* | 49.3 | *29.7* |
| | 3h | 53.9 | *55.5* | 51.1 | *33.0* | 47.4 | *21.1* | 54.9 | *7.2* | 47.5 | *26.9* |
| | **6h** | 44.9 | *55.9* | 41.9 | *25.6* | 40.3 | *36.4* | 43.2 | *13.1* | 41.2 | *40.3* |
| | 24h | 44.4 | *29.1* | 49.1 | *31.8* | 49.9 | *26.2* | 52.3 | *0.8* | 48.9 | *38.0* |
| | 27h | 35.0 | *8.3* | 53.2 | *23.6* | 57.2 | *34.9* | 58.5 | *8.7* | 45.7 | *24.8* |
| | 30h | 37.8 | *8.1* | 47.4 | *29.8* | 51.5 | *31.8* | 52.0 | *10.6* | 46.7 | *44.2* |
| | 48h | 41.0 | *24.1* | 45.2 | *19.9* | 49.4 | *34.1* | 42.2 | *26.4* | 49.4 | *34.6* |
| | 54h | 44.0 | *38.0* | 48.7 | *23.4* | 51.2 | *23.6* | 44.4 | *18.3* | 44.9 | *27.4* |
| | Median *IQR* | 43.8 | *32.9* | 48.6 | *26.7* | 49.8 | *28.4* | **52.5** | ***11.6*** | 47.6 | *33.5* |
| **Lactate** | Baseline | 1.4 | *2.1* | 1.4 | *1.1* | 1.6 | *0.6* | 1.8 | *0.9* | 1.7 | *0.8* |
| | 1h | 1.4 | *2.3* | 1.6 | *1.8* | 1.6 | *0.8* | 2.0 | *0.6* | 1.7 | *0.7* |
| | **3h** | 1.8 | *2.9* | 2.2 | *2.4* | 2.0 | *0.6* | 3.1 | *1.2* | 2.6 | *1.4* |
| | **6h** | 2.0 | *3.0* | 2.2 | *2.4* | 2.4 | *0.6* | 3.9 | *1.4* | 3.6 | *2.1* |
| | 24h | 1.3 | *0.9* | 1.3 | *0.8* | 1.2 | *0.7* | 2.2 | *0.6* | 1.8 | *1.5* |
| | 27h | 3.0 | *0.8* | 1.5 | *0.6* | 1.1 | *1.5* | 2.3 | *0.6* | 1.7 | *1.8* |
| | 30h | 2.8 | *0.3* | 1.4 | *1.4* | 1.0 | *0.8* | 2.0 | *0.6* | 1.7 | *2.0* |
| | 48h | 1.4 | *1.9* | 1.3 | *0.9* | 1.6 | *1.1* | 2.0 | *0.4* | 1.6 | *1.0* |
| | 54h | 1.5 | *1.7* | 1.3 | *1.0* | 1.5 | *1.2* | 2.0 | *0.3* | 1.7 | *0.9* |
| | Median *IQR* | 1.9 | *1.8* | 1.6 | *1.4* | 1.6 | *0.9* | 2.4 | *0.7* | 2.0 | *1.4* |
| **Glucose** | Baseline | 14.0 | *7.5* | 17.5 | *9.4* | 20.0 | *14.6* | 19.0 | *10.5* | 17.5 | *6.4* |
| | 1h | 14.0 | *7.5* | 16.5 | *10.5* | 21.0 | *16.5* | 19.0 | *4.5* | 16.0 | *8.6* |
| | 3h | 15.5 | *6.8* | 15.5 | *9.0* | 21.0 | *7.5* | 18.0 | *4.5* | 13.5 | *12.4* |
| | 6h | 15.5 | *8.6* | 17.0 | *9.0* | 25.0 | *13.5* | 18.0 | *4.5* | 17.5 | *7.1* |
| | 24h | 15.0 | *6.0* | 17.0 | *7.1* | 17.0 | *10.1* | 24.0 | *4.5* | 18.0 | *17.6* |
| | 27h | 22.5 | *12.8* | 16.0 | *5.6* | 17.0 | *6.0* | 21.0 | *9.0* | 19.5 | *16.1* |
| | 30h | 22.5 | *8.3* | 17.0 | *9.8* | 19.0 | *7.5* | 21.5 | *7.9* | 20.0 | *16.1* |
| | 48h | 16.0 | *10.9* | 17.0 | *4.9* | 21.5 | *11.3* | 23.0 | *16.5* | 18.0 | *9.8* |
| | 54h | 20.0 | *21.8* | 17.0 | *8.3* | 21.5 | *6.0* | 23.0 | *12.0* | 18.0 | *8.6* |
| | Median *IQR* | 17.2 | *10.0* | 16.7 | *8.2* | **20.3** | ***10.3*** | **20.7** | ***8.2*** | 17.6 | *11.4* |
| **BE** | Baseline | 3.7 | *5.9* | 2.8 | *3.5* | 4.3 | *0.7* | 4.6 | *2.4* | 4.1 | *2.2* |
| | 1h | 3.2 | *5.0* | 2.8 | *2.9* | 4.0 | *1.7* | 2.9 | *2.7* | 3.3 | *2.1* |
| | **3h** | 3.0 | *4.6* | 1.8 | *3.4* | 2.9 | *2.0* | 2.5 | *2.5* | 2.2 | *3.1* |
| | 6h | 3.2 | *6.9* | 2.9 | *4.0* | 4.0 | *2.7* | 3.3 | *2.6* | 3.9 | *3.0* |
| | 24h | 3.9 | *9.0* | 3.2 | *2.1* | 5.3 | *1.6* | 4.4 | *4.5* | 4.9 | *1.9* |
| | 27h | 3.7 | *6.3* | 2.1 | *1.5* | 3.2 | *3.8* | 3.8 | *3.4* | 4.5 | *3.4* |
| | 30h | 4.9 | *2.6* | 2.6 | *1.4* | 4.8 | *1.6* | 3.3 | *2.3* | 3.8 | *2.6* |
| | 48h | 3.2 | *5.2* | 1.7 | *4.4* | 4.5 | *3.9* | 2.9 | *2.4* | 2.9 | *3.1* |
| | **54h** | 2.9 | *4.9* | 1.6 | *3.9* | 4.4 | *2.0* | 3.8 | *2.0* | 2.8 | *3.1* |
| | Median *IQR* | 3.5 | *5.6* | **2.4** | ***3.0*** | 4.1 | *2.2* | 3.5 | *2.8* | 3.6 | *2.7* |

**Table 2. Cardiovascular responses** during LPS-induced fetal inflammatory response and vagus nerve manipulation (Median and interquartile range, IQR)**.**

| | | | | Experimental group | | | | | | | |
|---|---|---|---|---|---|---|---|---|---|---|---|
| **Parameter** | **Time Point** | **Control** | | **LPS** | | **Vx_LPS_400** | | **Vx_LPS_800** | | **Efferent VNS** | |
| **FHR** | Baseline | 165 | *34* | 155 | *29* | 149 | *24* | 148 | *21* | 159 | *27* |
| bpm | 1h | 153 | *51* | 155 | *39* | 146 | *26* | 161 | *32* | 155 | *32* |
| | 3h | 150 | *37* | 156 | *24* | 158 | *8* | 156 | *44* | 152 | *20* |
| | **6h** | 162 | *47* | 179 | *49* | 178 | *26* | 169 | *16* | 174 | *12* |
| | 24h | 158 | *75* | 158 | *43* | 139 | *22* | 147 | *43* | 172 | *45* |
| | 27h | 186 | *0* | 162 | *71* | 144 | *26* | 150 | *29* | 160 | *27* |
| | 30h | | | 167 | *26* | 147 | *29* | 155 | *40* | 160 | *35* |
| | **48h** | 161 | *18* | 168 | *46* | 151 | *39* | 161 | *48* | 172 | *30* |
| | 54h | 135 | *53* | 153 | *45* | 153 | *27* | 167 | *37* | 160 | *33* |
| | Median *IQR* | 159 | *39* | 161 | *41* | 152 | *25* | 157 | *34* | 163 | *29* |
| **dBP** | Baseline | 37 | *14* | 39 | *6* | 37 | *6* | 32 | *18* | 40 | *9* |
| mmHg | 1h | 38 | *18* | 37 | *13* | 39 | *2* | 39 | *31* | 41 | *18* |
| | 3h | 40 | *15* | 36 | *8* | 39 | *6* | 37 | *1* | 36 | *13* |
| | **6h** | 32 | *18* | 33 | *9* | 34 | *6* | 32 | *14* | 32 | *9* |
| | 24h | 39 | *12* | 34 | *11* | 41 | *8* | 34 | *6* | 37 | *12* |
| | 27h | 45 | *0* | 31 | *4* | 40 | *12* | 37 | *6* | 41 | *10* |
| | **30h** | | | 31 | *8* | 40 | *5* | 35 | *18* | 33 | *9* |
| | 48h | 35 | *8* | 35 | *6* | 31 | *5* | 45 | *7* | 34 | *10* |
| | 54h | 36 | *7* | 34 | *10* | 36 | *8* | 49 | *2* | 36 | *14* |
| | Median *IQR* | 38 | *11* | **34** | ***8*** | 37 | *6* | 38 | *12* | 37 | *12* |
| **mBP** | Baseline | 46 | *14* | 44 | *5* | 44 | *4* | 42 | *16* | 47 | *11* |
| mmHg | 1h | 43 | *20* | 44 | *11* | 46 | *7* | 44 | *33* | 46 | *15* |
| | 3h | 45 | *10* | 43 | *11* | 47 | *14* | 42 | *8* | 45 | *12* |
| | **6h** | 41 | *16* | 40 | *9* | 43 | *14* | 40 | *11* | 41 | *14* |
| | 24h | 46 | *11* | 41 | *15* | 46 | *6* | 42 | *10* | 44 | *7* |
| | 27h | 51 | *0* | 39 | *6* | 47 | *15* | 44 | *9* | 44 | *5* |
| | 30h | | | 38 | *10* | 46 | *12* | 43 | *16* | 45 | *8* |
| | 48h | 45 | *6* | 41 | *6* | 40 | *2* | 53 | *12* | 42 | *11* |
| | 54h | 43 | *4* | 39 | *6* | 44 | *9* | 59 | *1* | 42 | *8* |
| | Median *IQR* | 45 | *10* | 41 | *9* | 45 | *9* | 45 | *13* | 44 | *10* |
| **sBP** | Baseline | 54 | *15* | 50 | *9* | 49 | *17* | 54 | *12* | 58 | *16* |
| mmHg | 1h | 52 | *19* | 51 | *14* | 54 | *6* | 53 | *33* | 54 | *14* |
| | 3h | 48 | *12* | 51 | *13* | 55 | *3* | 50 | *11* | 51 | *13* |
| | **6h** | 51 | *13* | 45 | *12* | 53 | *24* | 47 | *9* | 51 | *18* |
| | 24h | 50 | *8* | 49 | *14* | 51 | *6* | 52 | *16* | 51 | *5* |
| | 27h | 57 | *0* | 45 | *13* | 55 | *18* | 51 | *11* | 50 | *8* |
| | 30h | | | 45 | *13* | 53 | *20* | 52 | *13* | 54 | *12* |
| | 48h | 56 | *27* | 46 | *12* | 49 | *2* | 62 | *22* | 52 | *13* |
| | 54h | 48 | *5* | 45 | *7* | 53 | *7* | 70 | *1* | 50 | *13* |
| | Median *IQR* | 52 | *13* | 47 | *12* | 52 | *12* | **54** | ***14*** | **52** | ***12*** |

FHR, fetal heart rate, beats per minute (bpm)
dBP, diastolic fetal arterial blood pressure (mmHg)
mBP, mean fetal arterial blood pressure (mmHg)
sBP, systolic fetal arterial blood pressure (mmHg)

**Table 3. Integrative overview of study's findings (cf. visual abstract).**

| System | Modality | GLM *(vs. Control and baseline time point)* | | | | Correlation | Comment |
|---|---|---|---|---|---|---|---|
| | | LPS | Vx+LPS400 | Vx+LPS800 | Efferent VNS | | |
| | Blood gas | none | | | | | |
| General | Lactate | 1.5x ↑ at 3 & 6h | | | | | |
| | FHR | ↑ at 6h & 48h with habituation | | | | | |
| | dBP | ↓ 6 & 30h | | | | | |
| | mBP | ↓ at 6h | | | | | |
| Cardio-vascular | sBP | ↓ 6h | | ↑ | | | Mild sepsis without overt cardiovascular decompensation; note the habituation effect |
| | Glucose | | ~1.3x ↑ | | | Control > eff VNS > Vx | Vx triggers chronic rise in glucose levels, starting prior to and regardless of LPS; eff. VNS restores this |
| Meta-bolic | Insulin | | ↑ at 3h; Vx > eff VNS | | | | Vx causes transient hyperinsulinemia at IL-6 peak, but not at baseline; eff. VNS restores this in part; Vx disrupts glucose-insulin homeostasis |
| | **IL-6** (systemic arterial) | ↑ 3 & 6h | ↓ to control levels | ↑ at 3 & 6h; Vx > LPS > eff VNS | | none with insulin | Vx+LPS400 surprisingly anti-inflammatory; Vx brake on inflammatory response is LPS dose dependent and ablates the memory (habituation to the 2nd hit); VNS reduces the inflammatory response and may boost the memory effect. |
| Inflammation | **CD11c** (terminal ileum) | ↑ | | ↑ | | | Similar to systemic response, we see LPS dose dependent M1 inflammatory response in ileum, restored with eff. VNS; we could not validate insulin's anti-inflammatory role via direct correlations to systemic or regional inflammation |